\documentclass[twocolumn,pra,a4paper,nofootinbib,showpacs,aps,floatfix,superscriptaddress]{revtex4}
\usepackage{graphicx}
\usepackage{bm}
\usepackage{amsmath}
\newcommand{\ignore}[1]{}
\def\beq{\begin{equation}}
\def\eeq{\end{equation}}

\begin{document}
\title{Bose-Einstein condensates on slightly asymmetric double-well potentials}

\author{B. Juli\'a-D\'\i az}
\affiliation{Departament d'Estructura i Constituents de la Mat\`{e}ria,\\
Universitat de Barcelona, 08028 Barcelona, Spain}
\affiliation{Institut de Ci\`{e}ncies del Cosmos,
Universitat de Barcelona, E--08028 Barcelona, Spain}

\author{J. Martorell}
\affiliation{Departament d'Estructura i Constituents de la Mat\`{e}ria,\\
Universitat de Barcelona, 08028 Barcelona, Spain}

\author{A. Polls}
\affiliation{Departament d'Estructura i Constituents de la Mat\`{e}ria,\\
Universitat de Barcelona, 08028 Barcelona, Spain}
\affiliation{Institut de Ci\`{e}ncies del Cosmos,
Universitat de Barcelona, E--08028 Barcelona, Spain}

\date{\today}
\begin{abstract}
An analytical insight into the symmetry breaking 
mechanisms underlying the transition from Josephson 
to self-trapping regimes in Bose-Einstein condensates is 
presented. We obtain expressions for 
the ground state properties of the system of a 
gas of attractive bosons modelized by a two 
site Bose-Hubbard hamiltonian with an external bias. 
Simple formulas are found relating the appearance 
of fragmentation in the condensate with the large 
quantum fluctuations of the population imbalance
occurring in the transition from the Josephson to 
the self-trapped regime. 
\end{abstract}

\maketitle

\section{Introduction}
The dynamics of cold bosonic atoms in double-well potentials 
has deserved a great deal of attention in the last decade. See
~\cite{Leggett01,pitaevski2003} for a careful review of 
the early findings, and \cite{GO07} for a more recent update. 
Of particular interest here are the seminal works of 
Smerzi {\it et al.}~\cite{Sme97} and 
Milburn {\it et al.}~\cite{Mil97}. The latter managed 
to derive a simplified many-body hamiltonian with 
semiclassical predictions similar to those of~\cite{Sme97} 
but with the important advantage of allowing the 
study of the quantum fluctuations on top of the 
semiclassical quantum averages. In both cases, the 
most remarkable feature reported was the existence 
of a new phenomenon, macroscopic self-trapping, 
directly linked to the atom-atom  interaction. Most of 
the studies focused on the case of repulsive atom-atom 
interactions. However, more recently interest has grown 
on the properties of these systems when this interaction is made
attractive and in particular on the appearance of cat like 
ground states whose description goes beyond the usual 
semiclassical approximations~\cite{Jav99,jame05,ST08}. 
The structure of these ground states is determined by 
the ratio between interaction and tunneling strengths. 

For repulsive interactions, the Bose-Hubbard model has 
been widely used to study the transition between the 
Josephson and the self-trapped regimes~\cite{Mil97,RW98,Jav99,PK01,MP05}. It 
is also the natural choice for attractive interactions. 
Varying the parameters of the model allows to easily  
scan the properties of the system. The transition from 
the Josephson to the self-trapped regime is directly 
related to the properties of the spectrum of the 
system as we increase the atom-atom interactions. In a recent 
manuscript the strongly correlated nature of 
the ground state of the system (for 
attractive interactions) in the transition 
region has been described~\cite{ours10}. This 
state is similar to the cat-like states described 
in Ref.~\cite{cirac98,SC98} for the case of internal 
Josephson-like behavior. Both studies find a 
deep relation between the appearance of a 
bifurcation in the semi-classical description 
and the existence of strongly correlated, cat-like, 
ground states in the systems. Similar states also appear 
in the nucleation of vortices in BECs~\cite{dagnino09}. 
The presence of the bifurcation points to the need of a 
fully quantum description that goes beyond the usual 
semiclassical approach. However the exact solution of 
the Bose-Hubbard model involves numerical calculations 
that complicate the interpretation of the results. It 
is thus highly desirable to develop also simplified models that 
give analytical insights into the relevant physics. 

In this article we extend  the previous work of 
Javanainen {\it et al.}~\cite{Jav99}, 
J\"a\"askel\"ainen and Meystre~\cite{jame05}, 
and the more recent one of  Shchesnovich and 
Trippenbach~\cite{tripp08,ST08}. Starting from 
Bose-Hubbard hamiltonians, these authors obtain 
approximate, Schr\"odinger-like, equations 
for the dynamics of the Fock space 
amplitudes of a system of $N$ atoms on 
a double-well potential. We will first rederive 
these equations and, afterwards, will go one 
step further and obtain simple 
analytical solutions under certain premises. 
These are found to be in good 
agreement with the exact results and help to 
explain the delicate balances
involved in the transition between the two regimes. 

\section{Quantum models for the ground state}
The time dependent Schr\"odinger equation 
governing the evolution of the system of 
$N$ atoms reads, 
\begin{eqnarray}
\frac{i}{N} \partial_{\tau}|\Phi> = \hat{H} |\Phi>\, ,
\label{eq:sff1}
\end{eqnarray} 
with the (Bose-Hubbard) Hamiltonian written in 
reduced units as~\footnote{To ease the comparison 
with the work in~\cite{ST08} 
we will make use of their notation.}, 
\begin{eqnarray}
\hat{H} &\equiv& -\frac{1}{N} (a_1^{\dag} a_2 + a_2^{\dag} a_1) +
\frac{\varepsilon}{N} \hat{n}_1 
+ \frac{\gamma}{N^2} (\hat{n}_1^2+ \hat{n}_2^2) \, .
\label{eq:sf1}
\end{eqnarray}
Where $\hat{n}_i=a^\dagger_ia_i$, $N$ is the total 
number of atoms in the system and $a_i$ ($a^\dag_i$) 
is the annihilation (creation) operator for well 
$i$. The Fock state basis is written as $|n_1, n_2\rangle$ 
and has $N+1$ vectors, $\{ |0,N\rangle, \dots, |N,0\rangle \}$. 
The two parameters governing the dynamics are 
$\gamma$, which measures the ratio between the 
contact atom-atom interactions and the hopping 
strength, and $\varepsilon$ which is a
symmetry breaking bias, also divided by the 
hopping strength \footnote{To compare to our previous 
work~\cite{ours10}, lets note that, $\Lambda = - 2 \gamma$, 
and $\varepsilon_{{\rm Ref} 12}/J = \varepsilon/2$.}. 
For this study we take $\varepsilon<0$ 
which promotes $|1\rangle$. 

With this sign convention, $\gamma>0$ and $\gamma<0$ 
correspond to repulsive and attractive atom-atom 
contact interactions.  

The solution to (\ref{eq:sf1}) can be expanded 
in the Fock basis as, 
$
|\Phi(t)> = \sum_{k=0}^N \ c_k(t) \ |k, N-k> 
%\label{eq:sf2}
$
, leading to an equation for the time evolution of 
the coefficients $c_k$ of the form~\cite{ST08}:
\begin{eqnarray}
\frac{i}{N}  \frac{d}{d\tau} c_k = -(b_{k-1}c_{k-1} + b_k c_{k+1}) + a_k
c_k \, ,
\label{eq:coef1}
\end{eqnarray} 
with $b_k = \frac{1}{N} \sqrt{(k+1)(N-k)}$, 
and 
$a_k = \frac{\gamma}{N^2} [k^2+ (N-k)^2] + \frac{\varepsilon}{N}k$.
Now, it is useful to introduce  a new variable, $x=k/N$, and, 
assuming that $h \equiv 1/N$ is small, we define 
$\Psi(x) = c_k/\sqrt{N}$ and $b(x)=b_k$. The next step is to 
make $x$ a continuous variable~\footnote{Formally, 
we then have, $b_{k-1}=e^{-h \partial_x} b(x)$, 
$c_{k-1}=\sqrt{N} e^{-h \partial_x}\Psi(x)$, and 
$c_{k+1}=\sqrt{N} e^{ h \partial_x}\Psi(x)$.}. 
This leads to Eq.~(5) of~\cite{ST08}, 
\begin{eqnarray}
i h \partial_{\tau} \Psi(x) &=& - \left[ e^{-i {\hat p}} b_h(x) + b_h(x)
  e^{i{\hat p}} \right] \Psi(x) + a(x) \Psi(x)\,,\nonumber \\
\label{eq:gf1}
\end{eqnarray}
where  
${\hat p} \equiv -ih \partial_x$, 
and
\begin{eqnarray}
b_h(x) &=& [(x+h)(1-x)]^{1/2} \, , \nonumber \\
a(x) &=& \gamma[x^2+(1-x)^2] + \varepsilon x \, .
\end{eqnarray}

Eq.~(\ref{eq:gf1}) describes the time evolution of 
$\Psi(x)$, which is a continuous interpolation of the 
$c_k/\sqrt{N}$. This interpolation assumes a certain 
degree of smoothness for the $c_k$, so that for most $k$: sign$(c_{k+1}) =$
sign$( c_k )$. This is the case for 
the ground state with either repulsive or 
attractive interactions, provided the symmetric state 
is lower in energy than the antisymmetric one 
(our case)~\footnote{There are however other cases for which the $c_k$
  alternate in sign : $c_k c_{k+1}<0$, $\forall k$. One example is the mean 
field state $\Phi= [(1/\sqrt{2})(a^\dagger_1-a^\dagger_2) ]^N|{\rm vac}\rangle$ 
As in~\cite{ST08}, one way of dealing with 
this case is by introducing an auxiliary smooth 
function $\tilde\Psi(x)=e^{\imath \pi k}\Psi(x)$.}.

From the formally exact Eq.~(\ref{eq:gf1}), we first 
retain up to terms of order $h^2$, 
\begin{eqnarray}
e^{\pm \imath \hat{p}} \simeq 1 \pm  h \partial_x 
- \frac{1}{2} h^2\partial^2_{x}
\end{eqnarray}
 and
\begin{eqnarray}
b_h(x)\simeq b_0(x)+ h\partial_h b_h(x)|_{h=0}
+ (1/2)h^2 \partial^2_h b_h(x)|_{h=0} \, ,
\end{eqnarray} 
to get,  
\begin{eqnarray}
i h \partial_{\tau} \psi(x) &\simeq& 
-h^2  b_0(x) \psi''(x) - h^2 b_0'(x) \psi'(x) \nonumber\\
&+& [a(x)+V_2(x)]\psi(x) 
\end{eqnarray}
where
\begin{eqnarray}
V_2(x) &=& -2 b_0(x) + h (b_0'(x)-  2\partial_h b_h(x)|_{h=0}  )  \nonumber \\
&-& h^2 (\partial_h^2 b_h(x)|_{h=0} -\partial_h b_h'(x)|_{h=0} 
+ {1\over 2} b_0''(x)) \, . \nonumber \\
\end{eqnarray}
This equation is similar to Eq.~(13) of~\cite{ST08} 
if we retain only the $h^0$ term in $V_2(x)$. This 
approximation respects the symmetry and 
behaves well at the edges of the Fock space. Besides,
taking into account  
that, 
\begin{eqnarray}
{\hat p} \sqrt{x(1-x)} {\hat p} \psi 
= - h^2 \partial_x(b_0 \partial_x \psi) = 
-h^2 (b'_0 \psi' + b_0 \psi'') \, , \nonumber \\
\end{eqnarray}
 we arrive to  
\begin{eqnarray}
ih \partial_{\tau} \psi &=& + [{\hat p} \sqrt{x(1-x)} {\hat p} + a(x)+V_0(x)]
\psi(x) \, , \nonumber \\ 
\label{eq:gf2}
\end{eqnarray}
with $V_0(x) = - 2\sqrt{x(1-x)}$. 

\begin{figure}[tb]                         %f01
\includegraphics[width=0.45\textwidth]{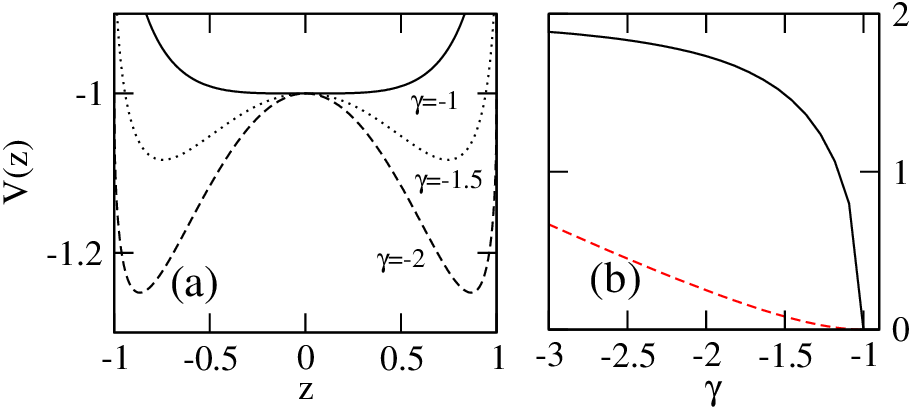}
\caption{(color online) (a) The potential ${\cal V}$ for several 
values of $\gamma$ and zero bias ($\epsilon =0$), 
(b) Separation between the two minima (solid-black) 
and height of the barrier as function of $\gamma$, 
(dashed-red), also in the case of zero bias. 
\label{fig:v}}
\end{figure}

The range of $x$ is the interval $[0,1]$. It will 
be convenient to introduce the variable 
$z\equiv 1- 2x$, $z\in [-1,1]$. In terms of the new variable, 
Eq.~(\ref{eq:gf2}) becomes
\begin{eqnarray}
i h \partial_{\tau} \psi(z) &=& \bigg( -2h^2 \partial_z \sqrt{1-z^2}
\partial_z  + {\cal V}(z) \bigg) \psi(z) 
\label{eq:sf5}
\end{eqnarray}
with ${\cal V}(z) = (1/2) (\gamma z^2 
-\varepsilon z) - \sqrt{1-z^2} \,.$
This equation is a Sch\"odinger-like equation 
with an effective $z$ dependent mass: $-h^2/(2 m ) \equiv -2h^2\sqrt{1-z^2}$. For 
convenience, we normalize $\psi(z)$ as 
$1 = \int_{-1}^1 \ dz \ |\psi(z)|^2$. 

As discussed in Refs.~\cite{tripp08} and ~\cite{ours10}, 
the population imbalance is an appropriate order 
parameter. It is defined as,  
\begin{eqnarray}
I = (\langle \Phi|\hat{n}_2- \hat{n}_1|\Phi\rangle)/N 
= \sum_{k=0}^N c_k^2 \left( 1 - 2 k/N \right) \, ,
\end{eqnarray}
which in the continuous variable reads, 
$I= \int_{-1}^1 \ dz |\psi(z)|^2 z\ ,$ 
where we have used $\sum_{k=0}^N c_k^2 = 1$ and 
the normalization of the $\psi(z)$. 

\subsection{Effective two mode model}
We first study the solutions of Eq.~(\ref{eq:sf5}) 
for the stationary case with zero bias, and concentrate 
on the region where the classical bifurcation appears, 
$\gamma < -1$. 

For $\gamma <-1$, ${\cal V}(z)$ has two symmetric 
minima, which get deeper as $\gamma$ decreases, 
see panel (a) of Fig.~\ref{fig:v}. The separation between 
the two minima, $2\sqrt{1-1/\gamma^2}$ and the 
barrier height ($\Delta=(1+\gamma)^2/(2|\gamma|)$) 
are depicted in panel (b).

\begin{figure}[t]                         %f01
\includegraphics[width=0.4\textwidth]{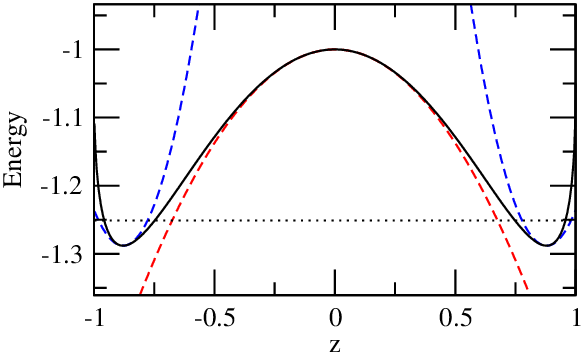}
\caption{(color online) The potential ${\cal V}$, black, compared to the 
three parabola used to approximate it near its extrema 
in building the analytical model. 
$N=50$, $\gamma=-2.1$.
\label{potpar}}
\end{figure} 

\begin{figure} [tb]                         %f01
\includegraphics[width=0.40\textwidth]{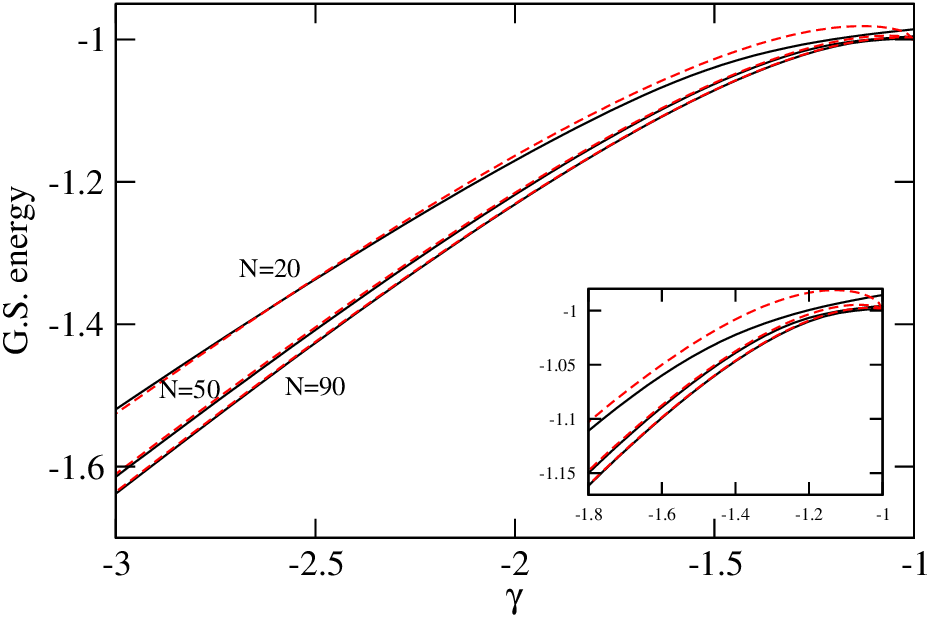}
\caption{(color online) Ground state energies of Eq.~(\ref{eq:sf5}), solid-black, 
compared to the approximate ones given by Eq.~(\ref{eq:t18}), 
dashed-red, for  $N = $ 20, 50 and 90, respectively. 
\label{fig:gse}}
\end{figure}

Since ${\cal V}(z)$ has  pronounced minima and 
maximum we first determine the g.s. energy of 
the system by approximating these minima by 
parabolas, see Fig.~\ref{potpar}, 
\begin{eqnarray}
{\cal V}(z)\simeq {\cal V}(z_m) + \frac{1}{2} {\cal V}''(z_m) (z \pm z_m)^2 
\end{eqnarray}
with $z_m = \sqrt{1-1/\gamma^2}$, 
${\cal V}(z_m) = \gamma/2+ 1/(2\gamma)$, and 
${\cal V}''(z_m)= \gamma- \gamma^3$ .
As an additional approximation, in the first term in the r.h.s. of
Eq.~(\ref{eq:sf5}) we replace $z$ by $z_m$, since the 
wavefunction, $\psi(z)$, is sharply 
peaked around the minima at $\pm z_m$. The stationary 
pseudo-Schr\"odinger equation can then be written as, 
\begin{eqnarray}
&&\left[-2h^2 \partial^2_z 
+ \frac{1}{2}(\gamma^4-\gamma^2) (z-z_m)^2\right]\psi\nonumber \\
&&= \left[-\gamma E +\left(1 +1/{\gamma^2} \right)/2
\right]\psi
\label{eq:t17}
\end{eqnarray}
and its spectrum can be readily obtained by 
comparing to the harmonic oscillator: 
$\hbar^2/(2m)\to 2h^2$, $m\omega^2 \to \gamma^2(\gamma^2-1)$ 
and $\hbar\omega/2 \to -h \gamma \sqrt{\gamma^2-1}$. From this 
analogy one easily gets the ground state energy:  
\begin{eqnarray}
 E_{\rm g.s.}&=& \frac{1}{2\gamma}+ \frac{\gamma}{2} + h \sqrt{\gamma^2-1} \ .
\label{eq:t18}
\end{eqnarray}  
This approximation turns out to be very reasonable as 
it can be seen in Fig.~\ref{fig:gse} by comparing the 
g.s. energies obtained using Eq.~(\ref{eq:t18}) and 
the ones obtained by solving numerically the 
pseudo-Schr\"odinger Eq.~(\ref{eq:sf5}). The 
approximation is particularly good for large 
$N$ ( small $ h =1/N$), and $\gamma$ not too close 
to the  bifurcation point, $\gamma=-1$, as shown 
in the inset of the figure. 

\begin{figure} [tb]                         %f01
\includegraphics[width=0.4\textwidth]{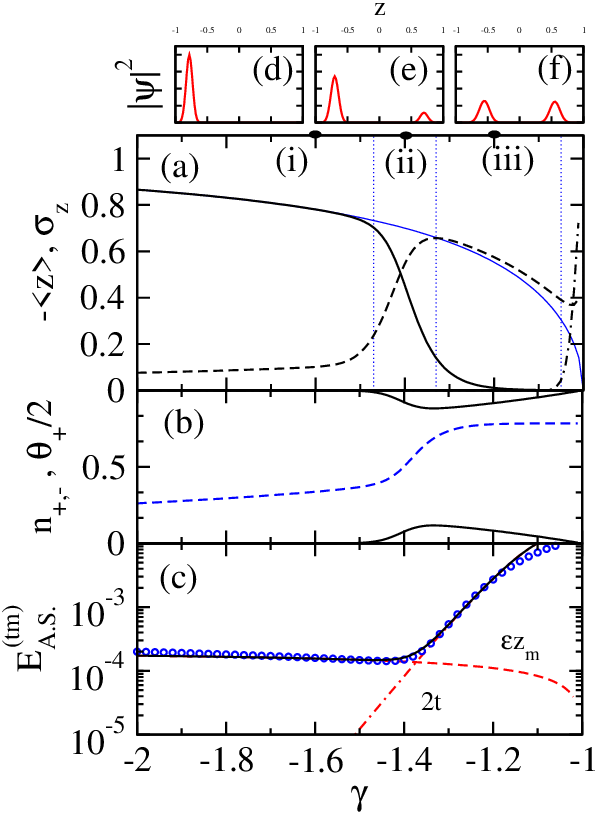}
\caption{(color online) The three smaller plots above, (d), 
(e), and (f), show $|\psi(z)|^2$
computed with the two mode model for $\gamma=-1.6, -1.4,$ 
and $-1.2$ respectively. (a) Different ground state 
properties predicted by the analytic model for three 
different regimes, separated by blue dotted lines. The 
three regions correspond to the bias-dominated 
fully-condensed (i), transition-asymmetric 
cat-like (ii), and symmetric cat-like (iii). 
We depict the behavior of $-\langle z\rangle_{g.s.}$, 
black solid line, and $\sigma_z$, black dashed line, 
as a function of $\gamma$. $z_m$ is shown in thin-blue. 
The overlap $\langle {\cal L}|{\cal R}\rangle$ is 
the black dot-dashed line. The middle panel, (b), 
contains the condensed fractions, $n_\pm$ (upper and 
lower black solid lines), and the mixing angle 
$\theta/2$ of the $\varphi_+$ eigenvector, see 
Eq.~(\ref{eq:dm3}), of the one-body density matrix 
associated to the ground state of the two-mode model. 
The lower panel, (c), depicts the energy splitting 
between the symmetric and antisymmetric (ground 
and first excited) states. 
The solid black line is calculated with 
Eq.~(\ref{eq:f23}). The blue empty circles are the 
exact numerical result obtained by solving 
Eq.~(\ref{eq:sf5}). Red dot-dashed and red 
dashed lines depict $2t$ and $\varepsilon z_m$ 
respectively. In all cases $\varepsilon=-0.0002$ 
and $N=50$. 
\label{fig3}}
\end{figure}

For our  analysis we  assume that 
the harmonic oscillator estimate of the ground state 
is accurate enough  and 
that the shape of $\psi(z)$ around $z_m$ (and also 
$-z_m$ ) is that of the g.s. harmonic oscillator, 
i.e. gaussian with the parameters also determined 
from Eq.~(\ref{eq:t18}). These normalized solutions 
will be denoted 
 $|{\cal R}\rangle$ 
or $\Psi_{\cal R}^{(h.o.)}(z)$, and $|{\cal L}\rangle$ or 
$\Psi_{\cal L}^{(h.o.)}(z)$. They correspond to the 
clockwise and counter-clockwise rotor solutions 
to the pendulum in the language of Ref.~\cite{Sme97}. 
They form the ``two-mode'' basis for the 
model to be described below. Notice also that these two modes are built in the
Fock space, they are useful to describe the system during the bifurcation and 
in the self-trapping regime.
In fact, the model works soon after the bifurcation 
starts, once the overlap $\langle {\cal L}|{\cal R}\rangle$ is 
small enough. This overlap is given by, 
$\exp [2 (\gamma^2-1)^{3/2}/(\gamma h)]$. 
As can be seen in panel (a) of Fig.~\ref{fig3}, this overlap 
(black dot-dashed line) is essentially zero for 
$\gamma \lesssim -1.05$ and $N=50$. For these basis vectors,  
$\langle {\cal L}|z|{\cal L}\rangle = -z_m$ 
and $\langle {\cal  R}|z|{\cal R}\rangle = z_m$.  
In this two-mode model, an approximate solution can be written as:  
\begin{eqnarray}
|\psi\rangle = \cos \alpha |{\cal L}\rangle + \sin \alpha |{\cal R}\rangle\,.
\label{eq:gs}
\end{eqnarray}

In the absence of bias the ground state is the symmetric 
combination, and therefore $\alpha =\pi/4$. The expectation 
value of $z$ in the  state (\ref{eq:gs}) is easily calculated, 
$
\langle z\rangle = -z_m (\cos^2 \alpha - \sin^2 \alpha) = 
-z_m \cos ( 2 \alpha)$. The dispersion of $z$, 
$\sigma_z^2\equiv \langle \psi|z^2|\psi\rangle 
-\langle \psi|z|\psi\rangle^2$,  can also be readily computed. 
Retaining up to order $h$,  
\begin{eqnarray}
\sigma_z^2 
= z_m^2\sin^2 2\alpha -h/(\gamma \sqrt{\gamma^2 - 1}) \, .
\end{eqnarray}
When no bias is present, $\langle z \rangle = 0$ and 
$\sigma_z^2 = z_m^2 -h/(\gamma \sqrt{\gamma^2 - 1})$.  
The large values of $\sigma_z^2$ are due to the fact 
that the ground state is cat-like, that is, the wavefunction has 
two peaks in $z$. 

When a finite but small bias, $|\varepsilon| <<1$, is considered, 
$\alpha$ is determined by the balance between the 
tunneling across the middle barrier of ${\cal V}$ 
and the bias term. In the absence of tunneling, 
$|\psi\rangle$ will consist only of $|{\cal L}\rangle$ or 
of $|{\cal R}\rangle$ depending on the sign of the bias 
constant, $\varepsilon$. In the latter situation the main 
contributor to $\sigma_z$ is the otherwise small 
$-h/(\gamma \sqrt{\gamma^2 - 1})$ term, as the $\sin^2 2\alpha$ 
becomes negligible. 

To take into account both effects, the tunneling across 
the Fock space barrier and the bias term, we write an 
effective Hamiltonian for this two-mode model
\begin{equation}
{\cal H} = 
\begin{pmatrix}   
\varepsilon z_m/2 & -t                \cr 
-t              & -\varepsilon z_m/2
\end{pmatrix}\,.
\label{eq:sf12b}
\end{equation}
The eigenvalues of ${\cal H}$ are 
$E_{S} =  -\sqrt{t^2+\varepsilon^2 z_m^2/4}=-E_A$. Thus, one 
way to obtain the value of the tunneling constant, $t>0$, is 
by computing the energy splitting between the symmetric 
and anti-symmetric states in the double well in the 
absence of bias, $t=(1/2)\Delta E_{AS}\equiv (1/2)(E_A-E_S)$. 

To estimate this energy splitting in the double-well 
we will neglect the $z$ dependence of the effective 
mass. This brings our problem to a Schr\"odinger-like 
equation and we can make use of the WKB 
approximation as developed in Razavy's book,~\cite{Ra03}. 
The energy splitting is written in Eq.~(3.109) of 
that book:
\begin{eqnarray}
\Delta E_{AS} = \frac{\hbar \omega}{\pi}  {\rm exp}\left(- \int_{z_-}^{z_+}
  \sqrt{(2m/\hbar^2)({\cal V}(z)-E) } dz\right)\, . \nonumber \\
\end{eqnarray} 
In our case we use the equivalences given above
Eq.~(\ref{eq:t18}) to assign values to $\hbar^2/(2m)$ and  
$\hbar \omega$. The value of $E$ corresponds to the ground-state energy 
and is taken from the harmonic oscillator approximation defined  in
Eq.~(\ref{eq:t18}).  
To compute the tunneling integral, and 
obtain analytic results, we will approximate 
${\cal V}(z)$ by an inverted parabola, 
${\cal V}(z) \simeq -1 + \frac{1}{2} (\gamma+ 1) z^2 $, see Fig.~\ref{potpar}. This 
gives an energy splitting, 
\begin{eqnarray}
\Delta E_{AS} &=& - 2 \frac{h \gamma}{\pi} \sqrt{\gamma^2-1}\; {\rm exp}\left( -
  \frac{\pi}{2h} \frac{|E|- 1}{ \sqrt{|\gamma|-1}}\right)\,.
\label{eq:raz3}
\end{eqnarray}
This estimate of the splitting in absence of bias, is used 
to extract the value of $t$ and it is in very good 
quantitative agreement with the exact energy splitting 
obtained by solving Eq.~(\ref{eq:sf5}) in absence of bias 
or as will be discussed later, when the bias is 
not dominant. It is an important improvement on the 
estimate given in Ref.~\cite{ST08} (cf. Eq.~(31)). The 
value of $2t$ is plotted (red dot-dashed line) in panel (c) of Fig. ~\ref{fig3}. 

From the two mode Hamiltonian (Eq.~(\ref{eq:sf12b})) we find 
that the predicted ``full'' (including the bias) 
symmetric-antisymmetric energy splitting is
\begin{equation}
\Delta E^{(\rm tm)}_{AS} = 2\sqrt{\left(\varepsilon z_m/2\right)^2+ t^2} 
\label{eq:f23}
\end{equation}
with two well-defined limits, $ 2t$  
and 
$\varepsilon z_m$ (red dot-dashed and red dashed lines of panel (c) of Fig.~\ref{fig3}
respectively), depending on whether 
tunneling or bias is the dominant contribution. 
The ground state can be readily computed, 
${\cal H} \psi_{S} = E_{S} \psi_{\rm S}$, with 
$\psi_{\rm S}=(\cos \alpha, \sin\alpha)$ and 
$\tan \alpha = \xi +\sqrt{\xi^2+1}$, where, 
$\xi \equiv (\varepsilon  z_m/2t)$. 
 
The one body density matrix for the Bose-Hubbard 
Hamiltonian can be written as, 
\begin{equation}
\hat{\rho}={1\over N}
\begin{pmatrix} 
\langle a^\dagger_1 a_1\rangle  &  \langle a^\dagger_1 a_2 \rangle              \cr 
\langle a^\dagger_2 a_1 \rangle & \langle a^\dagger_2 a_2\rangle  
\end{pmatrix}\,,
\end{equation}
with Tr${\hat \rho}=1$ and $\langle f(z)\rangle = \int dz |\Psi(z)|^2$. 
In the large-N model we have, $a^\dagger_1 a_1 =(1+z)/2$, 
$a^\dagger_2 a_2 =(1-z)/2$, $a^\dagger_1 a_2 =a^\dagger_2 a_1 
=\sqrt{1-z^2}/2$. In the two mode model retaining 
up to order $h^1$ we get, 
\begin{equation}
{\hat \rho} = \frac{1}{2}\begin{pmatrix} 1+ z_m \cos 2\alpha &
  \sqrt{1-z_m^2}\cr \sqrt{1-z_m^2} & 1-z_m \cos 2 \alpha \end{pmatrix}
\label{eq:dm1}
\end{equation}
with eigenvalues, 
\begin{eqnarray}
n_{\pm} &=& \frac{1}{2}\left( 1 \pm \sqrt{1 - z_m^2 \sin^2(2\alpha)}\right) \nonumber \\
&=& 
 \frac{1}{2}\left( 1 \pm \sqrt{1 - \sigma_z^2}\right) + {\cal O}(h)\, .
\end{eqnarray}
The last equality directly relates the appearance 
of fragmentation and the quantum fluctuations of 
the population imbalance measured by $\sigma_z$, and holds
also  in the 
transition region. Denoting the corresponding 
eigenvectors by: 
$\varphi_\pm=(\cos \theta_\pm /2, \sin\theta_\pm/2 )$ one finds
\begin{eqnarray}
\tan \frac {\theta_\pm}{2} = - \frac {\bigg[z_m \cos 2\alpha \mp \sqrt{1-z_m^2
  \sin^2(2\alpha)} \bigg] }{ \sqrt{1-z_m^2}}  \, .
\label{eq:dm3}
\end{eqnarray}

\subsection{Results}
The quantitative predictions of the analytical 
model are presented in Fig.~\ref{fig3}. The model 
provides a simple and yet deep understanding 
of the problem. For $\gamma<-1.5$, region (i), 
the ground state of the system is completely 
asymmetric (see the left upper smaller plot of Fig.~\ref{fig3}),
 with a large population imbalance, i.e. it 
is localized on the well promoted by the 
bias ($|{\cal L}\rangle$ for $\varepsilon<0$). 
The location of the wells in the Fock space 
is measured by $z_m$. The blue thin line 
in panel (a) shows $z_m$, which is a decreasing 
function of $\gamma$ and reaches zero at the 
bifurcation point, $\gamma=-1$.  In this regime, 
the bias dominates completely the hamiltonian 
of Eq.~(\ref{eq:sf12b}), which therefore is 
essentially diagonal with eigenvectors $|{\cal L}\rangle$ 
and $|{\cal R}\rangle$. In these cases, $\sigma_z$ is small 
and given by the spread of the occupied mode. In this 
region, the system is fully condensate. Correspondingly, 
as shown in the panel (b), the eigenvalues of the 
one-body density matrix of the ground state, are 1 and 0. 
In the panel (c), it is clearly shown that
$2t$ is very small compared with $\varepsilon z_m$, 
and the splitting $E_{A.S.}^{(tm)}$ (Eq.~\ref{eq:f23}) 
in this region is given by $\varepsilon z_m$.

In region (ii) both the bias term $\varepsilon z_m$ 
and the tunneling matrix element, $t$, are of comparable size 
(see panel (c)). In that region the 
ground state is an asymmetric cat-like 
state (see the central upper smaller plot). The value 
of -$\langle z\rangle $ is decreasing and has large 
fluctuations, shown in the  region (ii) of panel (a), 
mostly due to having both modes populated simultaneously. 
The system is fragmented into two condensates, i.e the 
eigenvalues of the one-body density matrix are both
 different than zero, being $\varphi_+$ more populated, 
see  panel (b) of the figure. The 
macro-occupied state, $\varphi_+$, 
varies from close to $|1\rangle$ to 
$(1/\sqrt{2})(|1 \rangle+|2 \rangle)$ as 
can be seen from the behavior of the mixing angle 
$\theta_+/2$ (blue dashed line in panel (b)). 
 
The tunneling term, $t$, grows exponentially 
and is responsible for the abrupt decrease 
of $z$ as one crosses into the 
region, (iii). There, 
the two mode Hamiltonian is  
dominated by the tunneling matrix element, 
and the wave function becomes symmetric-cat-like,
as shown in the right upper smaller plot. The cat like nature of 
the state reflects in a small value of the imbalance and a sizeable 
$\sigma_z$. Notice that when $\gamma$ approaches -1, 
the two-mode approximation is not valid anymore, 
because the $\langle {\cal L} \mid {\cal R} \rangle$, 
shown by the  black-dot-dashed line in panel (a) 
is not longer close to zero.

The simplifications of the two mode 
model capture to a very large extent 
the features of the initial large-N 
model, Eq.~(\ref{eq:sf5}), as can be seen 
from the comparison of the exact splitting (blue empty circles) 
between the ground and first excited 
state and the two-mode one (black solid line), shown in the  last 
panel of the figure. 

\section{Conclusions}

For attractive interactions, we have discussed here the appearance and
properties of the cat states  in the transition from the tunneling to the
interaction dominated regimes. The semiclassical approach shows that at some
critical value of the ratio of atom-atom to tunneling energies, $\gamma$, a bifurcation appears. We have constructed an
effective ``two mode model'' built from the solutions corresponding
to the branches of the bifurcation. And have shown that it is quite successful 
in describing not only the average imbalances in the presence of bias but also the
quantum fluctuations. The model includes the main quantal effects since its
predictions are in good agreement with exact
solutions of the Bose Hubbard hamiltonian. 

Our starting point is an already developed large-N 
model~\cite{Jav99,ST08} which gives accurate 
numerical predictions for $N\gtrsim 30$.  We have managed to obtain analytic 
expressions valid on the transition region from 
self-trapped (bias-dominated) to the 
symmetric cat-like region (dominated by the tunneling term). The main feature is the sharp change of the 
population imbalance
as $\gamma$ is decreased. It is  due to 
the delicate interplay between the bias and 
an effective tunneling term with a sharp dependence on $\gamma$. 

The fact that Eq.~(\ref{eq:sf5}) captures most of the 
non-linear dynamics of the original Bose-Hubbard hamiltonian 
implies that the complex dynamics of systems governed by 
non-linear terms can be studied very conveniently with
pseudo-Schrodinger equations governing the dynamics in Fock 
space. In these equations the main effects of non-linearity 
appear through a barrier-like
potential. In this way, some of the dramatic consequences of the 
non-linearities are mapped into exponentially varying magnitudes in 
linear systems. We have exploited this advantage to introduce further
approximations and construct a very simple effective two-mode 
model to reproduce and explain the main properties of the transition.

The authors thank M. Lewenstein for a careful reading of 
the manuscript. B.J-D. is supported by a CPAN CSD 2007-0042 
contract. This work is also supported by Grants 
No. FIS2008-01661, and No. 2009SGR1289 from Generalitat 
de Catalunya.

\end{document}